\documentclass{use_for_SLACpub}

\def\EPJC{{\em Eur.\ Phys.\ J.} C}
\def\JHEP{\em J.\ High Energy Phys.}

\begin{document}

\title{\boldmath Exploring the Weak Phase $\gamma$ in 
$B^\pm\to\pi K$ Decays\unboldmath}

\author{M. Neubert}

\address{Stanford Linear Accelerator Center, Stanford University\\
Stanford, California 94309, USA\\
E-mail: neubert@slac.stanford.edu}

\maketitle

\abstracts{
Measurements of the rates for the hadronic decays $B^\pm\to\pi K$
along with the CP-averaged $B^\pm\to\pi^\pm\pi^0$ branching ratio 
can be used to bound and extract the weak phase $\gamma
=\mbox{arg}(V_{ub}^*)$. Using preliminary CLEO data, we obtain the 
bounds $|\gamma|>93^\circ$ at $1\sigma$, and $|\gamma|>71^\circ$ at 
90\% CL.}

\section{Introduction}

The main target of present and future $B$ factories is the study 
of CP violation. It will provide tests of the flavor sector of the 
Standard Model, which predicts that all CP violation results 
from a single complex phase in the quark mixing matrix. The 
determination of the sides and angles of the unitarity triangle 
plays a central role in this program.\cite{BaBar} The angle 
$\beta=\mbox{arg}(V_{td}^*)$ is accessible in a theoretically
clean way through the observation of the CP asymmetry in the decay 
$B\to J/\psi K_S$, and a first measurement yielding $\sin 2\beta
=0.79_{-0.44}^{+0.41}$ has just been reported by the CDF 
Collaboration.\cite{CDF}

Recently, there has been significant progress in the theoretical 
understanding of the hadronic decays $B\to\pi K$, and methods have 
been developed to extract information on $\gamma=\mbox{arg}(V_{ub}^*)$ 
from measurements of the rates for these processes. Here, we discuss 
the charged modes $B^\pm\to\pi K$, which are particularly clean from 
a theoretical point of view.\cite{us}$^-$\cite{me} For applications 
involving the neutral decay modes the reader is referred to the 
literature.\cite{FM,Robert}

\section{\boldmath Theory of $B^\pm\to\pi K$ decays\unboldmath}

The hadronic decays $B\to\pi K$ are mediated by a low-energy
effective weak Hamiltonian, whose operators allow for three
distinct classes of flavor topologies: QCD penguins, trees, and 
electroweak (EW) penguins. In the Standard Model, the weak couplings 
associated with these topologies are known. Data show that the QCD 
penguins dominate the decay amplitudes, whereas trees and EW 
penguins are subleading and of a similar strength. The theoretical 
description of the two charged modes $B^\pm\to\pi^\pm K^0$ and 
$B^\pm\to\pi^0 K^\pm$ exploits the fact that the amplitudes for 
these processes differ in a pure isospin amplitude, $A_{3/2}$, 
given by the matrix element of the isovector part of the effective 
Hamiltonian between $B$ and the $(\pi K)$ isospin eigenstate with 
$I=\frac 32$. Up to an overall strong-interaction phase $\phi$,
the parameters of this amplitude are determined in the limit of 
SU(3) flavor symmetry.\cite{us} SU(3)-breaking corrections can be 
calculated in the factorization approximation,\cite{Stech} so that 
theoretical uncertainties enter only at the level of nonfactorizable 
SU(3)-breaking corrections to a subleading decay amplitude. 

A convenient parametrization of the decay amplitudes is
\begin{eqnarray}
   {\cal A}(B^+\to\pi^+ K^0) &=& P \Big[ 1
    - \varepsilon_a\,e^{i\gamma} e^{i\eta} \Big] \,,\nonumber\\
   - \sqrt2\,{\cal A}(B^+\to\pi^0 K^+) &=& P \Big[ 1
    - \varepsilon_a\,e^{i\gamma} e^{i\eta} 
    - \varepsilon_{3/2}\,e^{i\phi}
    (e^{i\gamma} - \delta_{\rm EW}) \Big] \,,
\end{eqnarray}
where $P$ is the dominant penguin amplitude defined as the sum of 
all terms in the $B^+\to\pi^+ K^0$ amplitude not proportional to 
$e^{i\gamma}$, $\eta$ and $\phi$ are strong phases, and 
$\varepsilon_a$, $\varepsilon_{3/2}$ and $\delta_{\rm EW}$ are 
real hadronic parameters. From a naive quark-diagram analysis, 
one does not expect the $B^+\to\pi^+ K^0$ amplitude to receive a 
contribution from $b\to u$ tree topologies; however, such a 
contribution can be induced through final-state rescattering or 
annihilation contributions.\cite{Blok}$^-$\cite{At97} They are 
parametrized by $\varepsilon_a$. Model estimates indicate that 
$\varepsilon_a$ could be about 5--10\%. In the future, it will be 
possible to put upper and lower bounds on this parameter by comparing 
the CP-averaged branching ratios for the decays $B^\pm\to\pi^\pm K^0$ 
and $B^\pm\to K^\pm\bar K^0$.\cite{Fa97} Below, we assume 
$\varepsilon_a\le 0.1$; however, our results will be almost 
insensitive to this assumption.

The parameter $\delta_{\rm EW}=0.64\pm 0.15$ describes the ratio of 
EW penguin and tree contributions to the isospin amplitude $A_{3/2}$. 
In the SU(3) limit, it is calculable in terms of Standard Model 
parameters.\cite{us,Fl96} The main uncertainty comes from the present 
errors on $|V_{ub}|$ and the mass of the top quark. SU(3)-breaking 
effects, which are included in the factorization approximation, 
reduce the value of $\delta_{\rm EW}$ by 6\%. In the error analysis 
we have assigned a 100\% uncertainty to this estimate. Note that, if 
nonfactorizable SU(3) breaking would lead to a further reduction of 
$\delta_{\rm EW}$, this would {\em strengthen\/} the bound on $\gamma$ 
derived below.

The parameter $\varepsilon_{3/2}$ describes the strength of the
$\Delta I=1$ tree amplitude relative to the leading penguin 
amplitude. We define a related parameter $\bar\varepsilon_{3/2}$
by writing $\varepsilon_{3/2}=\bar\varepsilon_{3/2}
\sqrt{1-2\varepsilon_a\cos\eta\cos\gamma+\varepsilon_a^2}$, so
that the two quantities agree in the limit $\varepsilon_a\to 0$. 
In the SU(3) limit, this new parameter can be determined 
experimentally form the relation
\begin{equation}\label{eps}
   \bar\varepsilon_{3/2} = \sqrt2\,R_{\rm SU(3)}
   \left|\frac{V_{us}}{V_{ud}}\right| \left[
   \frac{\mbox{B}(B^\pm\to\pi^\pm\pi^0)}
        {\mbox{B}(B^\pm\to\pi^\pm K^0)} \right]^{1/2} \,.
\end{equation}
Factorizable SU(3)-breaking is accounted for by a factor 
$R_{\rm SU(3)}\approx f_K/f_\pi\approx 1.2$. This may overestimate 
the effect; however, reducing the value of $\bar\varepsilon_{3/2}$ 
would again {\em strenghten\/} the bound on $\gamma$. We thus feel 
comfortable working with a large positive SU(3) correction and 
assign an error of 50\% to it to account for nonfactorizable effects. 
Using preliminary data reported by the CLEO Collaboration \cite{CLEO} 
combined with some theoretical guidance then yields 
$\bar\varepsilon_{3/2}=0.24\pm 0.06$.\cite{us} With a precise 
measurement of the CP-averaged branching ratios entering (\ref{eps}), 
the uncertainty in this number could be reduced significantly.

\section{\boldmath Lower bound on $\gamma$\unboldmath}

\begin{figure}
\epsfxsize=6cm 
\epsfbox{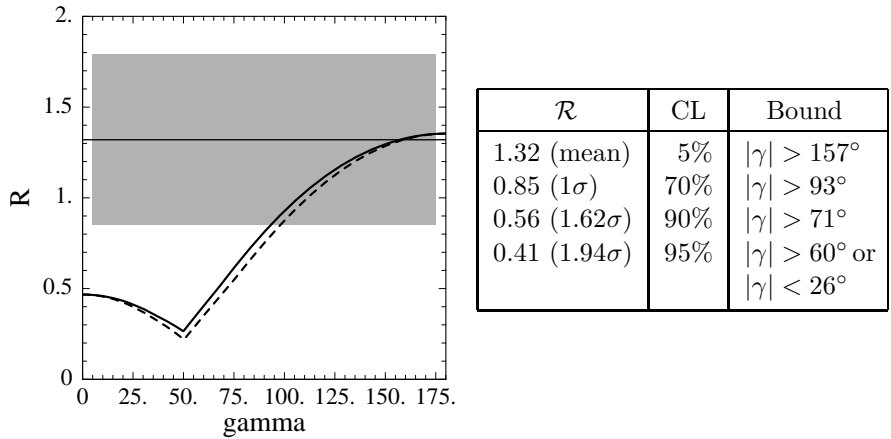} 
\raisebox{3.2cm}{
\begin{tabular}{|l|c|l|}
\hline
\qquad~\raisebox{0pt}[11pt][4.5pt]{${\cal R}$} & CL & ~~\,Bound \\
\hline
\raisebox{0pt}[11pt][0pt]{1.32} (mean) & \phantom{0}5\% &
 $|\gamma|>157^\circ$ \\
\raisebox{0pt}[9pt][0pt]{0.85} ($1\sigma$) & 70\% &
 $|\gamma|>93^\circ$ \\
\raisebox{0pt}[9pt][0pt]{0.56} ($1.62\sigma$) & 90\% &
 $|\gamma|>71^\circ$ \\
\raisebox{0pt}[9pt][0pt]{0.41} ($1.94\sigma$) & 95\% &
 $|\gamma|>60^\circ\,\mbox{or}\!$ \\
\raisebox{0pt}[9pt][6pt]{} & & $|\gamma|<26^\circ$ \\
\hline
\end{tabular}}
\caption{
Left: Theoretical upper bound on the ratio ${\cal R}$ versus 
$|\gamma|$ for $\varepsilon_a=0.1$ (solid) and $\varepsilon_a=0$ 
(dashed). The horizontal line and band show the current experimental 
value with its $1\sigma$ variation. Right: Bounds on $|\gamma|$ 
obtained for specific values of ${\cal R}$.
\label{fig:bound}}  
\end{figure}

Generalizing an idea of Fleischer and Mannel,\cite{FM} a bound on 
$\cos\gamma$ can be derived from a measurement of the ratio of the 
CP-averaged branching ratios for the two $B^\pm\to\pi K$ decay 
modes. The CLEO Collaboration has measured~\cite{CLEO}
\begin{equation}
   R_* = \frac{\mbox{B}(B^\pm\to\pi^\pm K^0)}
              {2\mbox{B}(B^\pm\to\pi^0 K^\pm)} 
   = 0.47\pm 0.24 \,.
\end{equation}
It is possible to derive an exact theoretical lower bound on $R_*$ 
by varying the strong phases $(\eta,\phi)$ independently between $-\pi$ 
and $\pi$. Knowing the other parameters ($\bar\varepsilon_{3/2}$
from data and $\delta_{\rm EW}$ from theory), this bound implies an
exclusion region for $\cos\gamma$, which becomes larger the smaller 
the values of $R_*$ and $\bar\varepsilon_{3/2}$ are. Since the 
$B^\pm\to\pi^\pm K^0$ branching ratio enters these two quantities 
in an opposite way, it is advantageous to consider the ratio 
${\cal R}=(1-\sqrt{R_*})/\bar\varepsilon_{3/2}$, for which the 
experimental error on this branching ratio tends to cancel.\cite{Frank} 
The current value is ${\cal R}=1.32\pm 0.47$. Theoretically, this ratio 
has the advantage of being independent of $\bar\varepsilon_{3/2}$ to 
leading order, and one obtains the bound~\cite{us}
\begin{equation}
   {\cal R} \le |\delta_{\rm EW}-\cos\gamma|
   + O(\bar\varepsilon_{3/2},\varepsilon_a) \,.
\end{equation}
An exact formula for the higher-order terms can be found in the
literature.\cite{me} For values ${\cal R}>0.8$ the exact bound is 
stronger than the approximate one shown above even after the 
rescattering effects parametrized by $\varepsilon_a$ are included. 
Varying the parameters in the intervals 
$0.18\le\bar\varepsilon_{3/2}\le 0.30$ and $0.49\le\delta_{\rm EW}
\le 0.79$, and setting $\varepsilon_a=0.1$, we obtain the bound 
shown in Figure~\ref{fig:bound}. Note that the effect of the 
rescattering contribution is very small. The table next to the 
figure shows the resulting bounds on $|\gamma|$ obtained at different 
confidence levels, which we obtained taking into account that in
the Standard Model the largest allowed value of ${\cal R}$ is 1.35.
(This is more conservative than assuming a two-sided Gaussian 
distribution.)

From a global analysis of the available information on the CKM 
matrix elements, one can derive indirect constraints on $\gamma$ 
that lead to the allowed range $37^\circ\le\gamma\le 118^\circ$, 
where the upper limit is determined by the current experimental limit 
on $B_s$--$\bar B_s$ mixing, whereas the lower bound is deduced from 
the measurement of CP violation in $K$--$\bar K$ mixing.\cite{BaBar} 
Without this information, i.e.\ using data from $B$ physics alone, 
this lower bound would disappear, and $\gamma=0$ would still be 
allowed. The 90\% CL bound on $|\gamma|$ derived here, combined with 
the upper bound from $B_s$--$\bar B_s$ mixing, implies that 
$71^\circ<|\gamma|<118^\circ$, which is a significant improvement 
over the range obtained from the global analysis. This, together with 
the fact that $|V_{ub}|\ne 0$ as shown by the existence of semileptonic 
$b\to u$ transitions, proves that the unitarity triangle is not 
degenerate to a line. In the context of the CKM model, this implies 
direct CP violation in $B$ decays.

\section{\boldmath Extraction of $\gamma$\unboldmath}

Ultimately, one would like not only to derive a bound on $\gamma$, 
but to measure this parameter directly from a study of CP violation. 
Recently, we have described a strategy for extracting $\gamma$ from 
$B^\pm\to\pi K$ decays,\cite{us2} which generalizes a method suggested 
by Gronau, Rosner and London~\cite{GRL} to include the effects of EW 
penguins. This approach has later been generalized to account for 
rescattering contributions to the $B^\pm\to\pi^\pm K^0$ decay 
amplitudes.\cite{me}

In addition to the ratio $R_*$, one considers the following 
combination of the direct CP asymmetries in the decays 
$B^\pm\to\pi K$:
\begin{equation}
   \widetilde A \equiv \frac{A_{\rm CP}(\pi^0 K^\pm)}{R_*}
   - A_{\rm CP}(\pi^\pm K^0) \,.
\end{equation}
With this definition, the rescattering effects parametrized by 
$\varepsilon_a$ are suppressed by a factor of $\bar\varepsilon_{3/2}$ 
and thus reduced to the percent level. The same is true for the ratio 
$R_*$. Explicitly, we have
\begin{eqnarray}
   R_*^{-1} &=& 1 + 2\bar\varepsilon_{3/2}\cos\phi\,
    (\delta_{\rm EW}-\cos\gamma) \nonumber\\
   &&\mbox{}+ \bar\varepsilon_{3/2}^2\,
    (1-2\delta_{\rm EW}\cos\gamma+\delta_{\rm EW}^2)
    + O(\bar\varepsilon_{3/2}\,\varepsilon_a) \,, \nonumber\\
   \widetilde A &=& 2\bar\varepsilon_{3/2} \sin\gamma \sin\phi
    + O(\bar\varepsilon_{3/2}\,\varepsilon_a) \,.
\end{eqnarray}
For fixed values of $\bar\varepsilon_{3/2}$ and $\delta_{\rm EW}$, 
these equations define contours in the $(\gamma,\phi)$ plane. When 
the rescattering corrections from $\varepsilon_a$ are included, 
these contours become narrow bands. From the intersection of the 
contour bands for $R_*$ and $\widetilde A$ one obtains the values of 
$\gamma$ and the strong phase $\phi$ up to possible discrete 
ambiguities. For some typical numerical examples, the theoretical 
uncertainties on the extracted values of $\gamma$ resulting from the 
variation of the input parameters $\bar\varepsilon_{3/2}$, 
$\delta_{\rm EW}$ and $\varepsilon_a$ are found to add up to a
total error of order $\delta\gamma\sim 10^\circ$.\cite{us2,me} A
precise determination of this error requires, however, to know the
actual values of $R_*$ and $\widetilde A$. For instance, Gronau
and Pirjol~\cite{Pirj} find a larger error for the specific case 
where the product $|\sin\gamma\sin\phi|$ is {\em very\/} close to 1, 
which would imply a value of the CP asymmetry $\widetilde A$ close 
to 50\%.

\section{Conclusions}

Measurements of the exclusive hadronic decays $B\to\pi K$ provide
interesting information on the weak phase $\gamma$. Using CLEO data 
on the CP-averaged branching ratios of the two charged decay modes, 
we have derived the bound $|\gamma|>71^\circ$ at 90\% CL. This bound
is stronger than the lower bound derived from the global analysis
of all other information on the CKM matrix. Combined with constraints
from $B_s$--$\bar B_s$ mixing and semileptonic $b\to u$ decays, and 
in the context of the CKM model, it proves the existence of direct 
CP violation in $B$ decays.

\newpage
\section*{Acknowledgments}
It is a pleasure to thank the organizers of WIN99, C.A.~Dominguez, 
R.D. Viollier and their staff, for arranging a marvellous meeting in 
a fantastic setting. I am grateful to Jon Rosner for collaboration 
on most of the work reported here. I also wish to thank Frank 
W\"urthwein for useful comments. This work was supported by the 
Department of Energy under contract DE--AC03--76SF00515.


\begin{thebibliography}{99}
\bibitem{Frank}
We are indebted to Frank W\"urthwein for pointing out the correlation
of errors in the measurements of $R_*$ and $\bar\varepsilon_{3/2}$, 
which was not taken into account in our original 
analysis.\protect\cite{us}
\end{thebibliography}
\end{document}